\documentclass[12pt,letterpaper]{article}
\usepackage[utf8]{inputenc}
\usepackage{amsmath}
\usepackage{booktabs}
\usepackage{tabu}
\usepackage{graphicx}
\usepackage{caption}
\usepackage{subcaption}
\usepackage{cite}
\usepackage{secdot}
\usepackage[margin=0.8in]{geometry}
\usepackage{amssymb}
\usepackage{hyperref}
\usepackage{ragged2e}
\usepackage[affil-it]{authblk}
\usepackage{abstract}

\usepackage[nodisplayskipstretch]{setspace} 
\begin{document}
\title{\Large{\textbf{Resonant leptogenesis and TM$_1$ mixing in minimal Type-I seesaw model with S$_4$ symmetry\vspace{-0.45em}}}}

\author{Bikash Thapa%
  \thanks{\texttt{bikash2@tezu.ernet.in} }
   \ and Ng. K. Francis
  \thanks{\texttt{francis@tezu.ernet.in}}}

\affil{\vspace{-1.05em}Department of Physics, Tezpur University, Tezpur - 784028, India}

\date{\vspace{-5ex}}
\maketitle
% \linenumbers
\begin{center}
\textbf{\large{Abstract}}\\
\justify
We present an S$_4$ flavour symmetric model within a minimal seesaw framework resulting in mass matrices that leads to TM$_1$ mixing. Minimal seesaw is realized by adding two right-handed neutrinos to the Standard Model. The model predicts Normal Hierarchy (NH) for neutrino masses. Using the constrained six-dimensional parameter space, we have evaluated the effective Majorana neutrino mass, which is the parameter of interest in neutrinoless double beta decay experiments. The possibility of explaining baryogenesis via resonant leptogenesis is also examined within the model. A non-zero, resonantly enhanced CP asymmetry generated from the decay of right-handed neutrinos at the TeV scale is studied, considering flavour effects. The evolution of lepton asymmetry is discussed by solving the set of Boltzmann equations numerically and obtain the value of baryon asymmetry to be $\lvert \eta_B \rvert  = 6.3 \times 10^{-10}$.
\end{center}

\hspace{-1.9em} PACS numbers: 12.60.-i, 14.60.Pq, 14.60.St
\newpage
\section{Introduction}
\label{sec:intro}
Neutrino oscillation experiments have determined that the mass of neutrinos are small but non-zero, and indicate flavour mixing \cite{particle2020review,aker2019improved,faessler2020status,araki2005measurement,PhysRevD.103.112010,doi:10.1142/S0217751X21300088}. Observations such as these pose a question about the origin of the tiny neutrino masses. The absence of the right-handed counterpart of the neutrinos within the standard model (SM) suggests that, unlike charged fermions, Dirac masses could not be the origin of neutrino masses. There are numerous frameworks beyond the standard model (BSM) that can explain the origin of neutrino masses, for instance, the seesaw mechanism \cite{minkowski1977mu, yanagida1980horizontal, mohapatra1986mechanism}, radiative seesaw mechanism \cite{ma2006verifiable},  models based on extra dimensions \cite{mohapatra1999neutrino,arkani2001neutrino}, and other models. The minimal seesaw mechanism, which is the extension of SM with two right-handed neutrinos, can explain the origin of neutrino masses as well as that of the Baryon Asymmetry of the Universe (BAU)  through leptogenesis \cite{fukugita1986barygenesis}. Here, the generation of baryon asymmetry involves conversion of lepton asymmetry, obtained from the CP-violating decay of the heavy right-handed neutrinos, via the sphaleron processes \cite{kuzmin1985anomalous}. It has been reported in Ref. \cite{davidson2002lower} that the mass scale $\mathcal{O}(10^9)$ for the right-handed neutrino is needed to explain the observed BAU. However, this scale can be lowered if we have nearly degenerate mass of right-handed neutrinos. Such a case leads to resonantly enhanced  CP-violating effects, and sufficient lepton asymmetry to account for BAU can be generated at relatively low masses (TeV scale). Such a situation is termed Resonant leptogenesis \cite{pilaftsis2004resonant}.\par
Further, the study of the origin of neutrino flavour mixing has received much attention through the years. Among various possibilities, tri-bimaximal mixing (TBM) \cite{harrison2002tri} seemed the most plausible explanation; however, experimental observations at Daya Bay \cite{an2012observation}, RENO \cite{ahn2012observation} and Double Chooz \cite{abe2012indication} suggest TBM requires corrections to incorporate $\theta_{13} \ne 0$. \iffalse There are several studies based on symmetries in the literature to explain the observed large mixing.\fi Trimaximal TM$_1$ \cite{luhn2013trimaximal,grimus2013discrete,rodejohann2012simple} is one such attractive mixing scheme that can be obtained by multiplying TBM with a 23-rotation leaving the first column of the TBM mixing matrix intact in TM$_1$ mixing, and the relationships among the observables in such a scheme are given as \cite{de2013flavour}
\begin{align}
\label{eq:1}
	 \sin^2 \theta_{12}  &= \frac{1-3 \sin^2 \theta_{13}}{3 \cos^2 \theta_{13}}, \\
\label{eq:2}
	 \cos \ \delta_{CP}  &= \frac{(1-5 \sin^2 \theta_{13})(2\sin^2 \theta_{23}-1)}{4\sin \theta_{13} \sin \theta_{23} \sqrt{2(1-3 \sin^2 \theta_{13})(1-\sin^2 \theta_{23})}}.
\end{align}
TM$_1$ mixing has proved to be compatible with the global data on neutrino oscillations, in a sense that it includes non-zero $\theta_{13}$ and agrees very well with the experiments on its prediction on the mixing angles $\theta_{13}$, $\theta_{23}$ and the Dirac phase, $\delta_{CP}$. Over the years, many discrete symmetry-based studies have been done  that gives rise to TM$_1$ mixing \cite{antusch2012trimaximal,de2013flavour,king2013minimal,chakraborty2020predictive,king2013lepton,king2016littlest}.
Herein, we propose a model based constructed using S$_4$ discrete symmetry within the framework of the minimal seesaw model. The resulting mass matrix leads to TM$_1$ mixing and can simultaneously explain BAU via resonant leptogenesis. The choice of right-handed neutrino Majorana mass matrix, M$_R$, is such that the right-handed neutrinos have degenerate mass at dimension five-level and successful resonant leptogenesis is achieved by introducing the higher-order term. In other words, our work is based on the extension of the model presented in \cite{zhao2011realizing}, such that it makes it suitable to study resonant leptogenesis within minimal seesaw scenario, and the orthogonality condition \cite{rodejohann2012simple, antusch2012trimaximal} allows us to realize TM$_1$ mixing in the leptonic sector.\par
\indent This paper is structured as follows. In section 2, we have presented the S$_4$ flavour symmetric minimal seesaw model followed by the features of the S$_4$ flavour group relevant to the construction of the model. Using the $3 \sigma$ range of neutrino oscillation data as constraints, we defined the allowed region for the model parameters in section 3. In section 4, the framework for resonant leptogenesis is described and the Boltzmann equations, which govern the evolution of the lepton number density and baryon asymmetry parameter, are numerically solved. It also includes the numerical results on neutrinoless double beta decay within the model and we finally conclude our work in section 5.

\section{Model Framework}
The S$_4$ flavour symmetry has been widely used to explain the observed flavour mixing of neutrinos \cite{luhn2013trimaximal,king2016littlest,zhao2011realizing,koide2007s4,krishnan2013simplest, brown1984neutrino,Bazzocchi2009FermionMA,bjorkeroth2017natural,meloni2010see,chen2020new}. S$_4$ group is a non-Abelian discrete group of permutations of 4 objects. It has 24 elements and 5 irreducible representations \textbf{$1_1$}, \textbf{$1_2$}, \textbf{$2$}, \textbf{$3_1$} and \textbf{$3_2$}. The product rules and Clebsch-Gordon coefficients are presented in Appendix A. In this work, we have considered the extension of the standard model (SM) with a discrete non-abelian group S$_4$. Also, a Z$_3 \times$Z$_2$ group is introduced to avoid specific unwanted couplings and achieve desired structures for the mass matrices. The fermion sector includes, in addition to the SM fermions, two right-handed neutrinos N$_1$ and N$_2$\iffalse which transforms as singlets 1$_1$ and 1$_2$ under S$_4$, respectively\fi. Flavons $\varphi_l$, $\phi_l$, $\varphi_\nu$, $\phi_\nu$, $\chi$, $\psi$, $\beta$ and $\rho$ forms the extension in the scalar sector. The charges carried by the various fields under different symmetry groups are presented in Table \ref{tab:1}. Following the representations of the fields given in Table \ref{tab:1}, we can write the invariant Yukawa Lagrangian
\begin{equation}
\label{eq:3}
\begin{split}
	-\mathcal{L} \  \supset \ & \frac{y_{l_1}}{\Lambda} \bar{L} H 				\varphi_l e_R + \frac{y_{l_2}}{\Lambda} \bar{L} H \varphi_l (\mu_R, \tau_R) + 	\frac{y_{l_3}}{\Lambda} \bar{L} H \phi_l (\mu_R, \tau_R)\\ & +  					\frac{y_{\nu_1}}{\Lambda}\bar{L} \tilde{H} \varphi_\nu N_1 + 						\frac{y_{\nu_2}}{\Lambda}\bar{L} \tilde{H} \phi_\nu N_2 + 						\frac{y_{\nu_3}}{\Lambda}\bar{L} \tilde{H} \psi N_2 \\ &+ y_{N_1} 						\bar{N_1^c} 	N_1 \beta + y_{N_2} \bar{N_2^c} N_2 \beta  + y_{N_3} \bar{N_1^c} 		N_2 \rho \frac{\beta \beta}{\Lambda^2} + h.c.,
\end{split}
\end{equation}
\noindent where $H$ is SM Higgs doublet and $\tilde{H} = i \sigma_2 H^*$, $\sigma_2$ being the 2$^\textrm{nd}$ Pauli matrices. The vacuum expectation values ($vev$) of the scalar fields are of the form \cite{zhao2011realizing}
\begin{equation}
\label{eq:4}
\begin{split}
	&\langle \varphi_l \rangle = (v_{\varphi_l}, 0, 0),\quad
	\langle \phi_l \rangle = (v_{\phi_l}, 0, 0), \quad
	\langle \varphi_\nu \rangle = (0, -v_{\varphi_\nu}, v_{\varphi_\nu}),\\ & 			\langle \phi_\nu \rangle = (v_{\phi_\nu}, v_{\phi_\nu}, v_{\phi_\nu}), 				\qquad \langle \beta \rangle = v_\beta, \quad \quad \langle \rho \rangle = v_			\rho.
\end{split}	
\end{equation} 
\noindent As for the $vev$ of $\psi$ we choose $\langle \psi \rangle = (0, -v_{\psi}, v_{\psi})$ following the orthogonality conditions $\langle \psi \rangle \cdot\langle \phi_l \rangle$ and $\langle \psi \rangle \cdot\langle \phi_\nu \rangle$. After electroweak and flavour symmetry breaking, we obtain the following structure for the charged lepton mass matrix
\begin{equation}
\label{eq:5}
	m_l = \frac{v_H}{\Lambda}
	\begin{pmatrix}
		y_{l_1} v_{\varphi_l} & 0 & 0 \\
		0 & y_{l_2} v_{\varphi_l} + y_{l_3} v_{\phi_l} \\
		0 & 0 & y_{l_2} v_{\varphi_l} - y_{l_3} v_{\phi_l}
	\end{pmatrix}.
\end{equation}
\renewcommand{\arraystretch}{1}
\begin{table}[t]
	
	\centering
	
	\begin{tabu}{ c c  c c c c c c c c c c c c c }
	
	\toprule
	Field & $\bar{L}$ & $e_R$ & ($\mu_R$, $\tau_R$) & $N_1$ & $N_2$ & $H$ & $					\varphi_l$ & $\phi_l$ & $\varphi_\nu$ & $\phi_\nu$ & $\psi$ & $\xi$ & $				\zeta$\\
	\toprule
	S$_4$ & 3$_1$ & 1$_1$ & 2 & 1$_1$ & 1$_2$ & 1$_1$ & 3$_1$ & 3$_2$ & 3$_1$ 			& 3$_2$ & 3$_2$ & 1$_1$ & 1$_2$\\
		
	Z$_3$ & 1 & $\omega^2$ & $\omega^2$ & 1 & 1 & 1&$\omega$ & $\omega$ & 1 & 1 
	& 1 & 1 & 1\\

	Z$_2$ & 1 & 1 & 1 & -1 & 1 & 1 & 1 & 1 & -1 & 1 & 1 & 1 & -1\\
	\bottomrule

\end{tabu}
	\caption{Field content and their representations under S$_4  \times $ Z$_3 \times $Z$_2$.}
	
	\label{tab:1}
\end{table}

The charged lepton sector of the model is similar to that of \cite{zhao2011realizing} and we similarly assume that the Froggatt-Nielsen mechanism explains the observed mass hierarchy of the charged leptons. Similarly, using the $vev$ presented in Eq.(\ref{eq:4}) for the neutrino sector, we obtain the Dirac and Majorana mass matrix
\begin{align}
\label{eq:6}
	m_D &= 
	\begin{pmatrix}
		0 & b\\
		a & b+c\\
		-a & b-c
	\end{pmatrix}
	,\\
\label{eq:7}
	m_R &= 
	\begin{pmatrix}
		M & 0\\
		0 & M
	\end{pmatrix},
\end{align}
\noindent where $a = y_{\nu_1} \frac{v_H v_{\varphi_\nu}}{\Lambda}$, $b = y_{\nu_2} \frac{v_H v_{\phi_\nu}}{\Lambda}$, $c = y_{\nu_3} \frac{v_H v_{\psi}}{\Lambda}$ with $v_H$ being the $vev$ of the SM Higgs. Taking $y_{N_1} \simeq y_{N_2} = y_N$ we have degenerate masses for the right-handed neutrinos, $M = y_{N} v_\xi$\footnote[1]{Here we obtain degenerate masses for the right-handed neutrino considering terms upto dimension-5}.\par
In the seesaw framework, the resultant light neutrino mass matrix is given by the well-known formula
\begin{align}
\label{eq:8}
	m_\nu &= -m_D m^{-1}_R m^T_D \\ 
\label{eq:9}
		&= \frac{1}{M}
	\begin{pmatrix}
		b^2 & b(b+c) & b(b-c)\\
		b(b+c) & a^2+(b+c)^2 & -(a^2-b^2+c^2)\\
		b(b-c) & -(a^2-b^2+c^2) & a^2+(b-c)^2
	\end{pmatrix}
	\\
\label{eq:10}
	&= \begin{pmatrix}
		b'^2 & b'(b'+c') & b'(b'-c')\\
		b'(b'+c') & a'^2+(b'+c')^2 & -(a'^2-b'^2+c'^2)\\
		b'(b'-c') & -(a'^2-b'^2+c'^2) & a'^2+(b'-c')^2
	\end{pmatrix},
\end{align}
\noindent with $a' = \frac{a}{\sqrt{M}}$, $b' = \frac{b}{\sqrt{M}}$ and $c' = \frac{c}{\sqrt{M}}$. In charged-lepton diagonal basis, the neutrino mixing matrix, $U_{\nu}$, is the unitary matrix that diagonalizes the mass matrix in Eq.(\ref{eq:10}). The resulting $U_\nu$ matrix, which is determined entirely from the neutrino sector, is 
\begin{equation}
\label{eq:11}
	U_\nu = U_{\textrm{TBM}} U_{23} = U_{\textrm{TM}_1},
\end{equation}
\noindent where $U_{\textrm{TBM}}$ is the tri-bimaximal mixing (TBM) matrix, $U_{23}$ is a unitary matrix whose (1,2), (1,3), (2,1), (3,1) entries are vanishing and the resulting matrix, $U_{\textrm{TM}_1}$, has its 1$^{\textrm{st}}$ column coinciding with that of TBM matrix.\\
The diagonalization equation thus reads 
\begin{equation}
\label{eq:12}
	U_\nu^T m_\nu U_\nu = diag(m_1, m_2, m_3).
\end{equation} 
The light neutrino masses are given as,
\begin{equation}
\label{eq:13}
	m_1 = 0, \quad
	m_2 = \frac{1}{2} |(s - \sqrt{t + s^2})|, \quad
	m_3 = \frac{1}{2} |(s + \sqrt{t + s^2})|,
\end{equation} 
\noindent where $s = 2 a'^2 + 3 b'^2 + 2 c'^2$ and $t = -24 a'^2 b'^2$. It is evident from Eq.(\ref{eq:13}) that the model predicts normal hierarchy of light neutrino.\par
In the following sections, we have presented the numerical approaches and discussed baryogenesis via resonant leptogenesis, neutrinoless double beta decay within the context of our model.

\section{Numerical Analysis}
In the previous section, we have shown how the S$_4$ model can be implemented in the minimal seesaw scenario, which results in mass matrices that lead to TM$_1$ mixing and normal hierarchy (NH) of masses for the neutrinos. Here, we perform numerical analysis to see the model's implication on leptogenesis and other phenomenological predictions. The mass matrix in Eq.(\ref{eq:10}) gives the effective neutrino mass matrix in terms of the complex model parameters $a'$, $b'$, and $c'$. We find the values of the model parameters by fitting the model to the current neutrino oscillation data. To do so, we use the $3 \sigma$ interval \cite{esteban2020fate} for the neutrino oscillation parameters ($\theta_{12}$, $\theta_{23}$, $\theta_{13}$, $\Delta m_{21}^2$, $\Delta m_{31}^2$) as presented in Table \ref{tab:2}. A further constraint on the model parameters was applied on the sum of absolute neutrino masses $\sum_i m_{i}<0.12$ eV from the cosmological bound \cite{aghanim2020planck}.

\begin{table}[t]
	\centering
	
	{\tabulinesep=1.3mm
	\begin{tabu}{ c c c }
	\toprule
	Parameters & Best-fit $\pm 1\sigma$ & $3\sigma$ range\\
	\toprule
	$\Delta m_{21}^2$[$10^{-5}$eV$^2$] &  $7.42^{+0.21}_{-0.20}$ & $6.82-8.04$\\
	
	$\Delta m_{31}^2$[$10^{-3}$eV$^2$] (NH) &  $2.517^{+0.026}_{-0.028}$ & 					$2.435-2.598$\\
	
	$\sin^2 \theta_{12}$ & $0.304^{+0.012}_{-0.012}$ & $0.269-0.343$\\
	
	$\sin^2 \theta_{23}$ & $0.573^{+0.016}_{-0.020}$ & $0.415-0.616$\\
	
	$\sin^2 \theta_{13}$ & $0.02219^{+0.00062}_{-0.00063}$ & $0.02032-0.02410$\\
	
	$\delta_{CP}/\pi $ (NH) & $1.09^{+0.15}_{-0.13}$ & $0.667-2.05$\\
	\bottomrule
	
	\end{tabu}}
	\caption{Neutrino oscillation parameters used to fit the model parameters.}
\label{tab:2}
\end{table}
\noindent In our analysis, the three complex parameters of the model are treated as free parameters and are allowed to run over the following ranges:
\begin{align*}
	|a'| \in [0.1, 0.2] \ \textrm{eV}^{1/2}\textrm{, }\quad |b'| &\in [0.03, 0.06] \ \textrm{eV}^{1/2}\textrm{, } \quad |c'| \in [10^{-4}, 0.1] \ \textrm{eV}^{1/2}\textrm{, }\\
	\phi_a \in [-\pi, \pi]\textrm{, }\qquad\qquad&\phi_b \in [-\pi, \pi]\textrm{, }\qquad\qquad \phi_c \in [-\pi, \pi],
\end{align*}
\noindent where $\phi_a$, $\phi_b$, $\phi_c$ are the phases given by $arg(a')$, $arg(b')$, $arg(c')$, respectively. Using relation $U^\dagger \mathcal{M} U = \textrm{diag}(m_1^2, m_2^2, m_3^2)$, with $\mathcal{M} = m_\nu m_\nu^\dagger$ and $U$ is a unitary matrix, we numerically diagonalize the effective neutrino mass matrix $m_\nu$. The mixing angles, $\theta_{23}$, $\theta_{13}$ are obtained using the relation 
\begin{equation}
\label{eq:14}
	\sin^2\theta_{23} = \frac{\lvert U_{23} \rvert^2}{1-\lvert U_{13} \rvert^2}, \quad	\sin^2\theta_{13} = \lvert U_{13} \rvert^2.
\end{equation}
\noindent As seen from Eqs.(\ref{eq:1}) and (\ref{eq:2}), TM$_1$  mixing gives correlations among the mixing angles and CP phases. These relations are assumed to calculate the observables $\theta_{12}$ and $\delta_{CP}$. The points in the 6-dimensional parameter space which corresponds to the observables that satisfy the $3\sigma$ bound on neutrino oscillations are taken to be the allowed region and the best-fit values for the model parameters ($\lvert a'\rvert$, $\lvert b'\rvert$, $\lvert c'\rvert$, $\phi_a$, $\phi_b$,$\phi_c$)  correspond to the minimum of the following $\chi^2$ function
\begin{equation}
	\label{eq:15}
	\chi^2 = \sum_{i}\left(\frac{\lambda_i^{model} - \lambda_i^{expt}}{\Delta \lambda_i}\right)^2,
\end{equation}
\noindent where $\lambda_i^{model}$ is the $i^{th}$ observable predicted by the model, $\lambda_i^{model}$ stands for the $i^{th}$ experimental best-fit value (Table \ref{tab:2}) and $\Delta \lambda_i$ is the 1$\sigma$ range of the $i^{th}$ observable.

\begin{figure}[!ht]
	\begin{center}
		\begin{subfigure}{0.485\textwidth}
			\includegraphics[width=\textwidth]{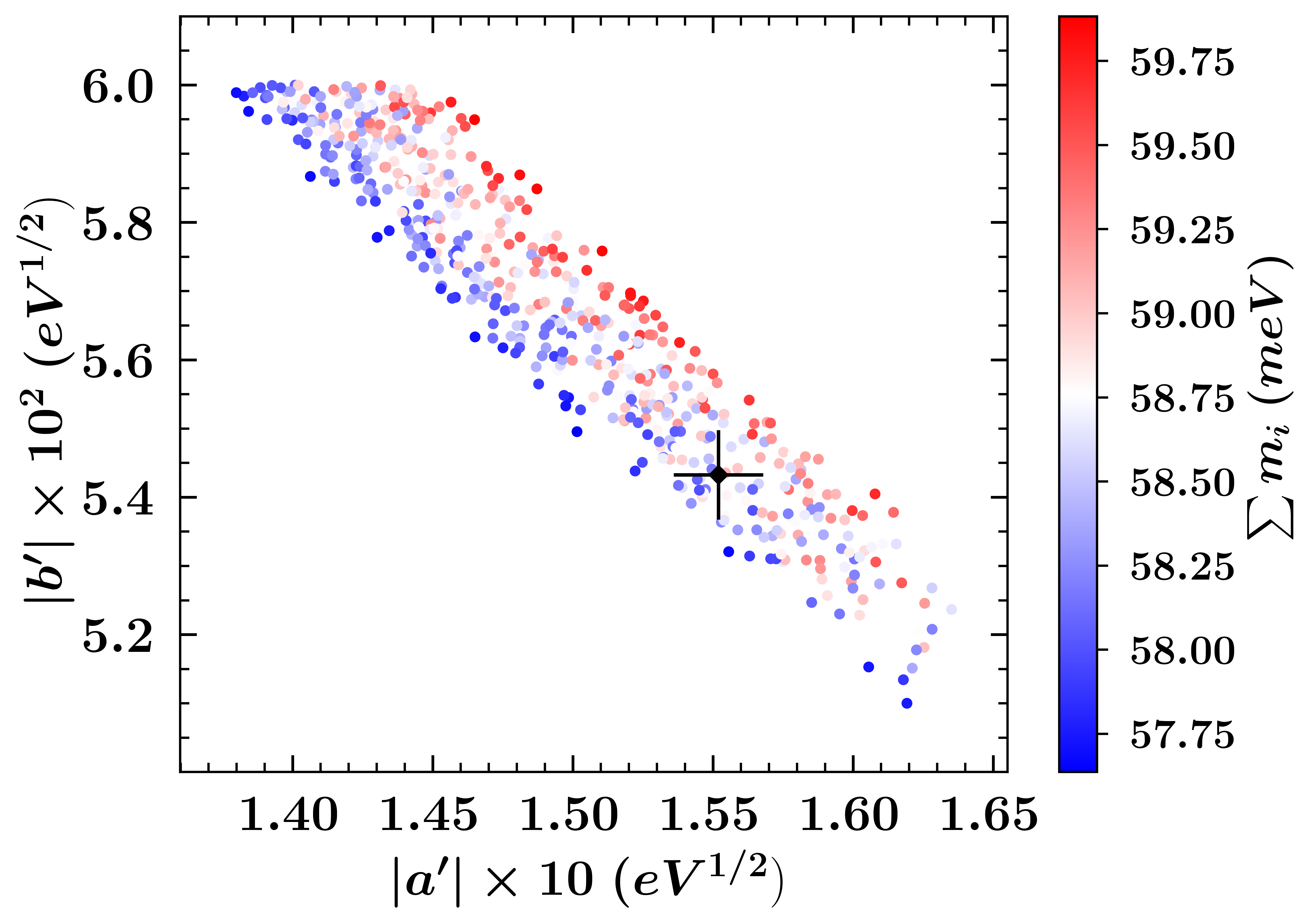}
		\end{subfigure}
		\hfill
		\begin{subfigure}{0.485\textwidth}
			\includegraphics[width=\textwidth]{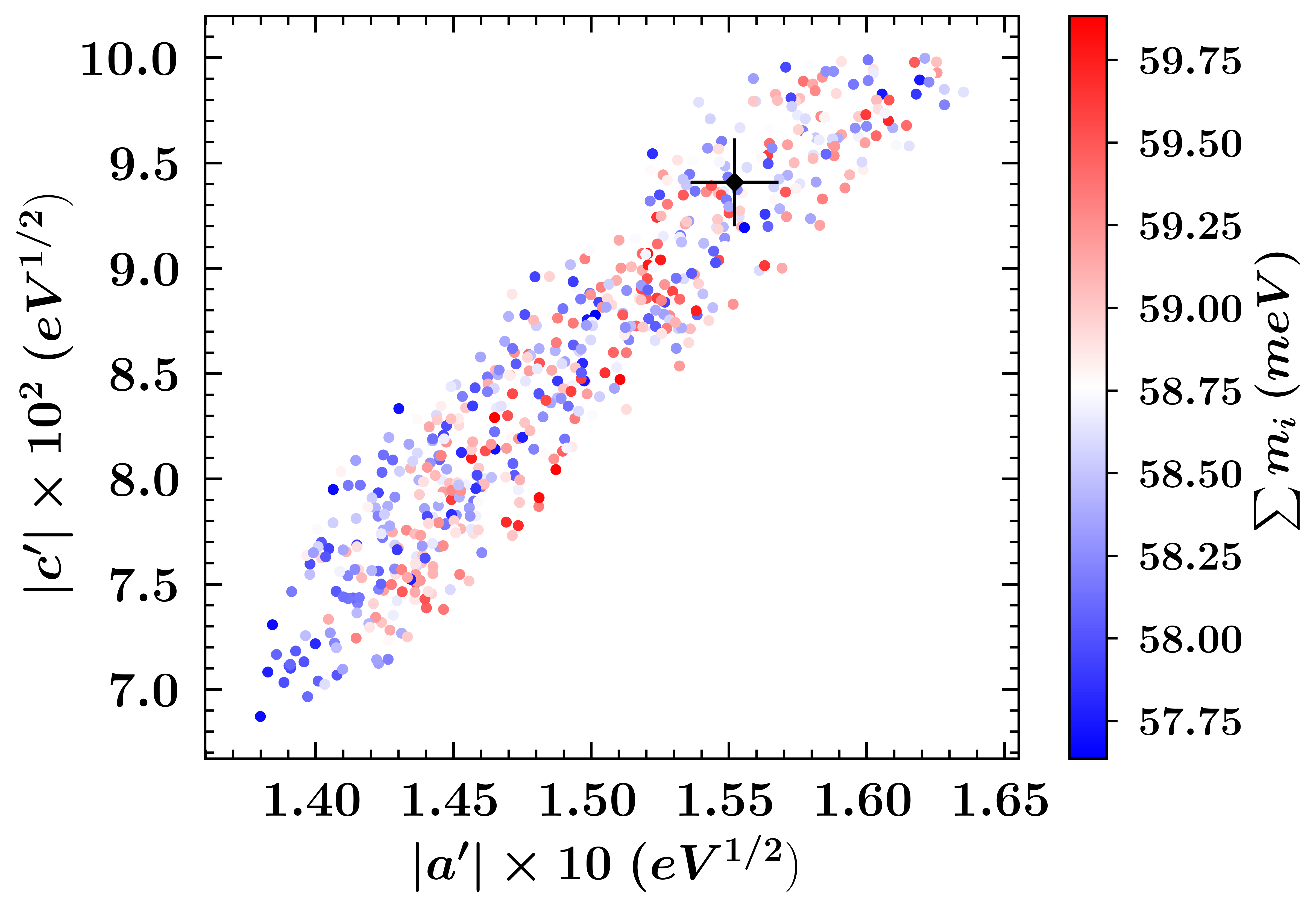}
		\end{subfigure}
		\caption{Left and right panel shows the correlation of $\lvert a'\rvert$ with $\lvert b'\rvert$ and $\lvert c'\rvert$ respectively along with the variation of $\sum m_i$. The cross mark indicate the best-fit values with $\chi^2 = \chi_{min}^2$, which corresponds to $\sum m_i = 0.0586$ eV.}
		\label{fig:mp1}
	\end{center}
\end{figure}

\begin{figure}[!ht]
	\begin{center}
		\begin{subfigure}{0.45\textwidth}
			\includegraphics[width=\textwidth]{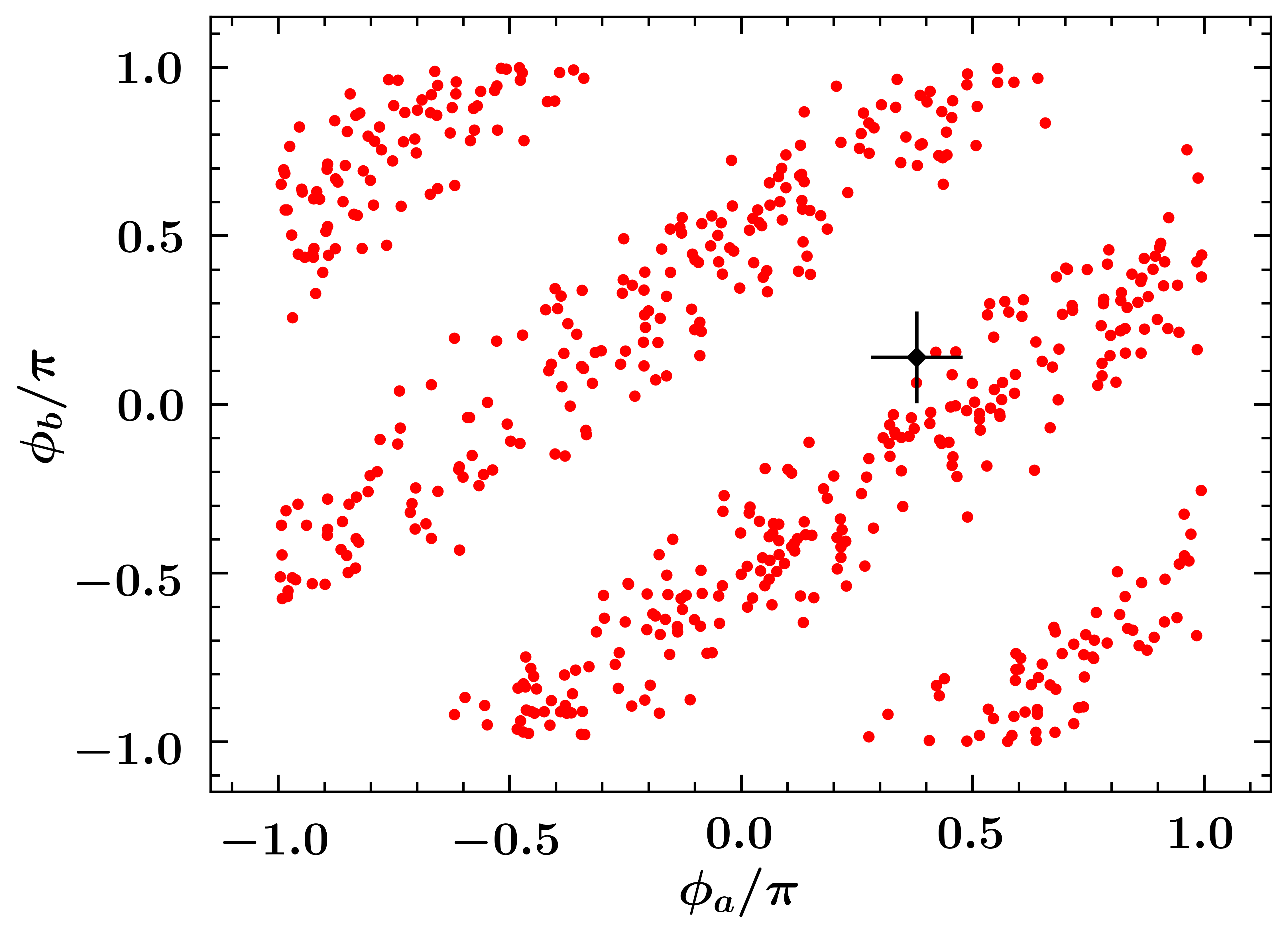}
		\end{subfigure}
		\hfill
		\begin{subfigure}{0.45\textwidth}
			\includegraphics[width=\textwidth]{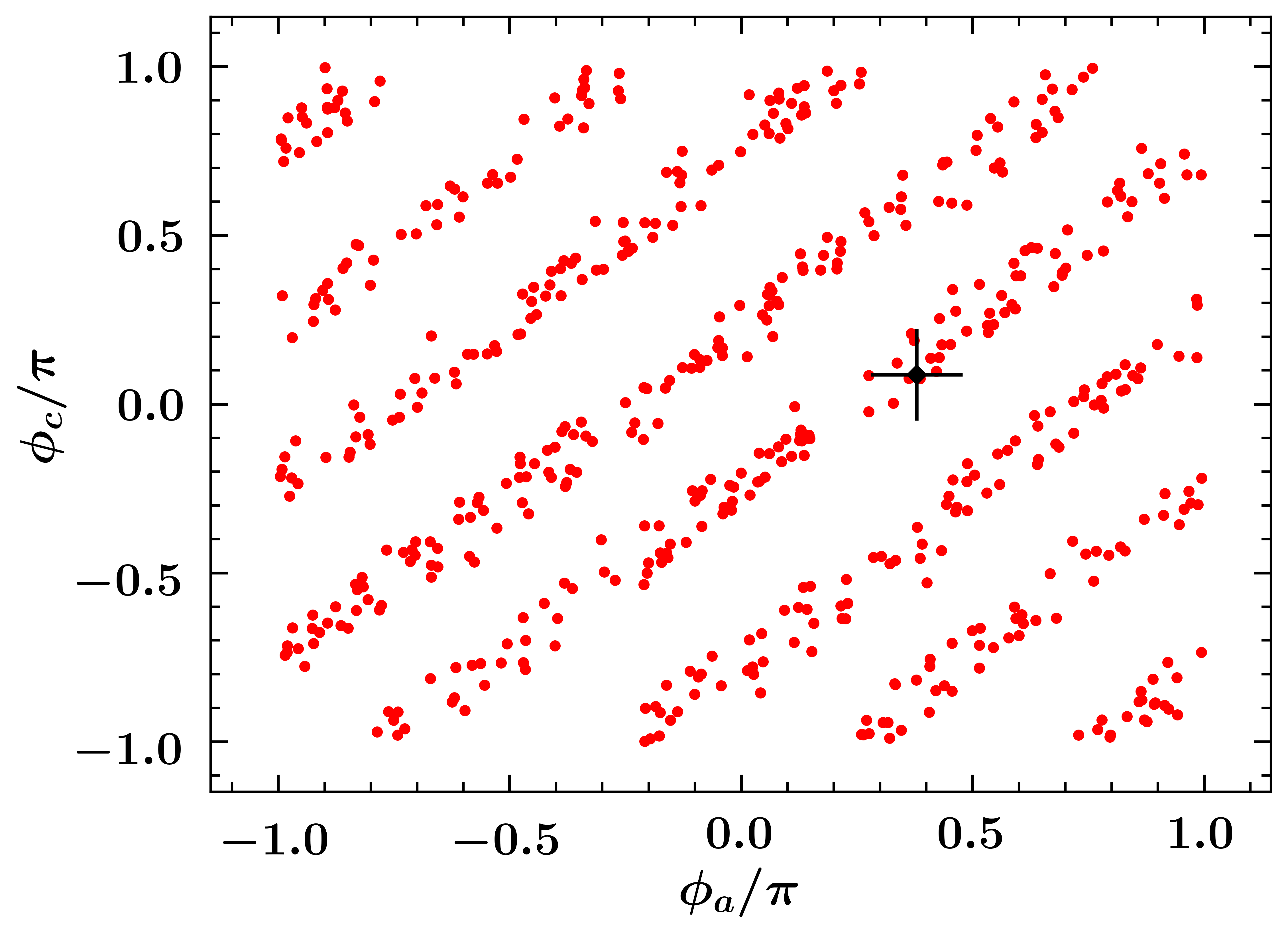}
		\end{subfigure}
		\caption{Correlation between the phases $\phi_a$, $\phi_b$ and $\phi_c$. The cross mark indicate the best-fit values corresponding to $\chi_{min}^2$.}
		\label{fig:mp2}
	\end{center}
\end{figure}

\vspace{1mm}

\par
\begin{figure}[!ht]
	\begin{center}
		\includegraphics[width=0.6\textwidth]{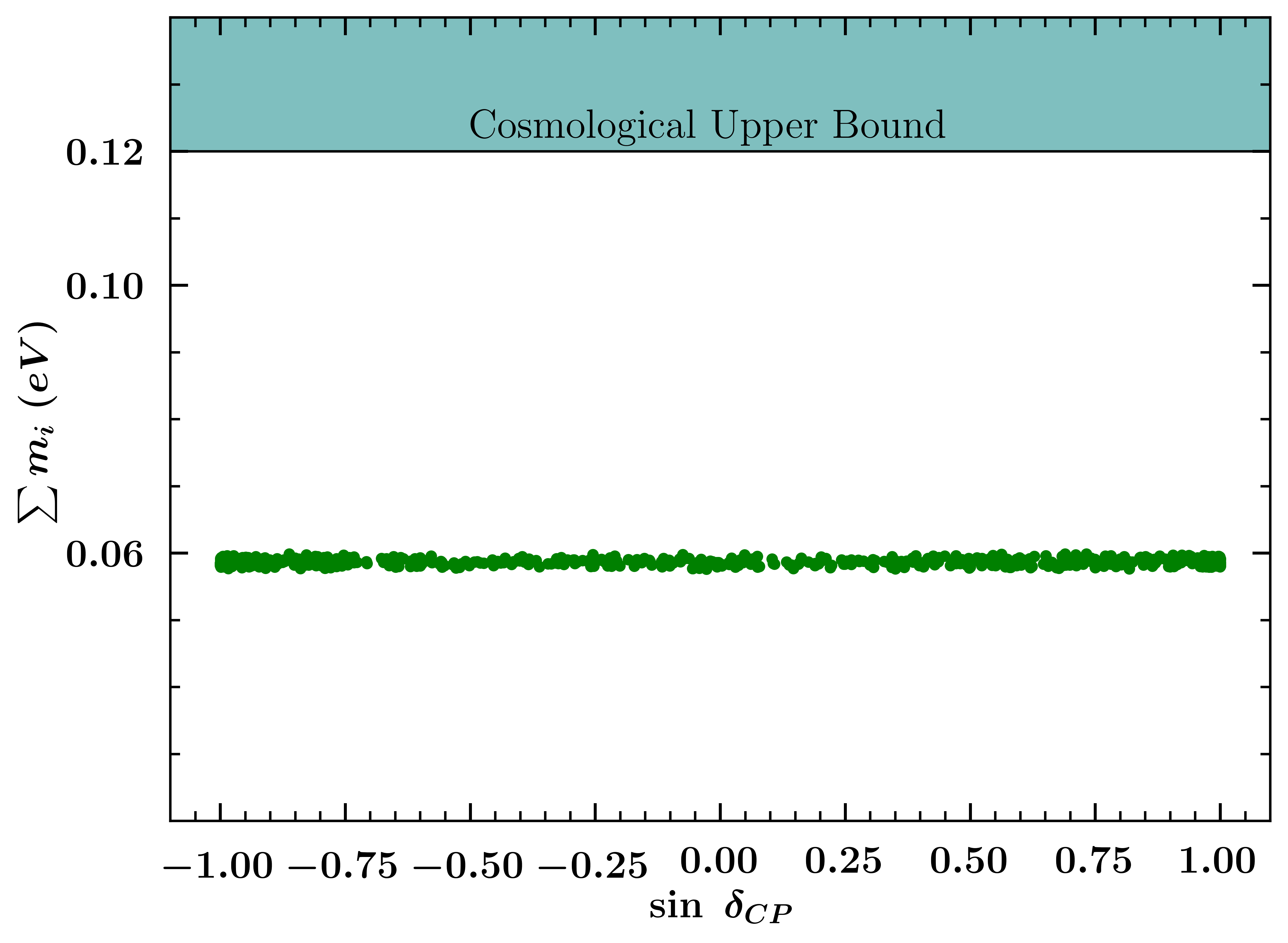}
		\caption{Shows the correlation of $\sum m_i$ with $\sin \delta_{CP}$ predicted in the model.}
		\label{fig:del}
	\end{center}
\end{figure}

The allowed regions for the various model parameters are given in Figure \ref{fig:mp1} and \ref{fig:mp2}. The best-fit values for $\lvert a'\rvert$, $\lvert b'\rvert$, $\lvert c'\rvert$, $\phi_a$, $\phi_b$ and $\phi_c$ obtained using the function defined in Eq.(\ref{eq:15}) are (0.155, 0.054, 0.09, 0.379$\pi$, 0.139$\pi$, 0.087$\pi$), denoted by the cross mark. Correspondingly, the best-fit values for the neutrino oscillation parameters are: $\sin^2 \theta_{12} = 0.318$, $\sin^2 \theta_{23} = 0.592$, $\sin^2 \theta_{13} = 0.02225$, $\sin \delta_{CP} = -0.454$, $\Delta m_{21}^2 = 7.25 \times 10^{-5}$ eV$^2$ and $\Delta m_{31}^2 = 2.51 \times 10^{-3}$ eV$^2$. We have shown the value of $\sin \delta_{CP}$ predicted by the model in Figure \ref{fig:del}.

\vspace{1mm}

\section{Resonant Leptogenesis}
The mechanism of leptogenesis, first proposed by Fukugita and Yanagida \cite{fukugita1986barygenesis}, is one of the popular mechanisms that can explain the observed baryon asymmetry of the universe (BAU). In the simplest scenario of thermal leptogenesis with a hierarchical mass spectrum of right-handed neutrinos, there is a lower bound on the mass of the lightest right-handed neutrino, M$_1 \simeq 10^9$ GeV \cite{davidson2002lower}. Although one can lower this limit if their masses are nearly degenerate, the scenario is popularly known as resonant leptogenesis \cite{pilaftsis2004resonant, pilaftsis1997cp}. In such a situation, one-loop self-energy contribution is enhanced resonantly, and the flavour-dependent asymmetry produced from the decay of right-handed neutrino into lepton and Higgs is given by \cite{xing2020bridging, pilaftsis1997resonant, anisimov2006cp, de2007resonant,francis2012validity}:
\begin{align}
\label{eq:16}
	\varepsilon_{i\alpha} &= \frac{\Gamma\left(N_i \rightarrow l_\alpha +H\right)\;-\;\Gamma\left(N_i \rightarrow \bar{l}_\alpha +\bar{H}\right)}{\sum_\alpha\Gamma\left(N_i \rightarrow l_\alpha +H\right)\;+\;\Gamma\left(N_i \rightarrow \bar{l}_\alpha +\bar{H}\right)}\\
	&=\sum_{i \neq j}\frac{\textrm{Im}\left[\left(Y_\nu^*\right)_{\alpha i} \left(Y_\nu^*\right)_{\alpha j}\left(Y_\nu^\dagger Y_\nu\right)_{ij} + \xi_{ij}\left(Y_\nu^*\right)_{\alpha i} \left(Y_\nu^*\right)_{\alpha j}\left(Y_\nu^\dagger Y_\nu\right)_{ji}\right]}{\left(Y_\nu^\dagger Y_\nu\right)_{ii} \left(Y_\nu^\dagger Y_\nu\right)_{jj}}\nonumber\\
\label{eq:17}
	& \hspace{7.5cm}\cdot \frac{\xi_{ij}\zeta_j\left(\xi_{ij}^2 - 1\right)}{\left(\xi_{ij}\zeta_j\right)^2 + \left(\xi_{ij}^2 - 1\right)^2}. 
\end{align}
\noindent where $\xi_{ij} = M_i/M_j$ and $\zeta_j = \left(Y_\nu^\dagger Y_\nu\right)_{jj}/(8\pi)$ with $Y_\nu = m_D/v$. 
\\
\noindent In our model, we have two right-handed neutrinos with exactly degenerate masses, M$_1 = $ M$_2 = $ M. However, successful leptogenesis requires a tiny mass splitting between the two right-handed neutrinos, which is introduced by adding a higher dimension term in the model (Eq.\ref{eq:3}). Such term leads to a minor correction in the Majorana mass matrix of Eq.(\ref{eq:7}), and the resultant structure of the mass matrix may be written as 
\begin{equation}
\label{eq:18}
	m_R = 
	\begin{pmatrix}
		M & \epsilon  \\
		\epsilon  & M
	\end{pmatrix},
\end{equation}
\noindent where $\epsilon = y_{N_3} v_\rho \frac{v_\beta^2}{\Lambda^2}$ is a parameter that quantifies the tiny difference between masses required for leptogenesis \footnote[2]{$\epsilon$ is assumed to be real.}. The mass matrix in Eq.(\ref{eq:18}) is diagonalized using a (2 $\times$ 2) matrix of the form
\begin{equation}
\label{eq:19}
	U_R = \frac{1}{\sqrt{2}}
	\begin{pmatrix}
		1 & 1 \\
		-1 & 1
	\end{pmatrix},
\end{equation} 
\noindent with real eigenvalues M$_1 = \textrm{M} - \epsilon$ and M$_2 = \textrm{M} + \epsilon$. In the basis where the charged-lepton and Majorana mass matrix are diagonal, the dirac mass matrix (Eq.\ref{eq:6}) takes the form
\begin{equation}
\label{eq:20}
	m'_D = \frac{1}{\sqrt{2}}
	\begin{pmatrix}
		-b & b\\
		a - (b+c) & a + (b+c)\\
		-a-(b-c) & -a+(b-c)
	\end{pmatrix}.
\end{equation}
\noindent From this point onward, we will take $Y_\nu = m'_D/v$, which is relevant for calculating CP asymmetry that arises during the decay of right-handed neutrinos in out-of-equilibrium way. Taking the best-fit values of the model parameters obtained in the previous section, we solve the following coupled Boltzmann equations describing the evolution, with respect to $z =\textrm{M}_1/T$, of RH neutrino density, $N_{N_i}$ and lepton number density for three flavours, $N_{\alpha\alpha}$ corresponding to  $\alpha = e,\ \mu,\ \tau$ \cite{pilaftsis2004resonant,de2007resonant}.
\begin{align}
\label{eq:21}
	\frac{dN_{N_i}}{dz} & = -D_i\ \left(N_{N_i}-N_{N_i}^{\textrm{eq}}\right)\\
	\frac{dN_{\alpha\alpha}}{dz} & = -\sum_{i=1}^2 \varepsilon_{i\alpha} D_i \left(N_{N_i}-N_{N_i}^{\textrm{eq}}\right)\nonumber \\  & \qquad -\frac{1}{4} \left\{\sum_{i=1}^2 (r z)^2 D_i  \mathcal{K}_2(r z) + W_{\Delta L=2}\right\} N_{\alpha\alpha}.
\end{align}
\noindent The following equation gives the equilibrium number density of N$_i$,
\begin{equation}
\label{eq:22}
N_{N_i}^\textrm{eq} = \frac{45 g_N}{4\pi^4 g_*}z^2 \mathcal{K}_2(z),
\end{equation}
\noindent with $\mathcal{K}_{1,2}(z)$ being the modified Bessel function. The parameter, $D_i$, sometimes called the decay parameter is defined as 
\begin{equation}
\label{eqn:23}
	D_i = \frac{z}{H(z=1)}\frac{\Gamma_{N_i}}{N_{N_i}^{\textrm{eq}}},
\end{equation} 
which gives the total decay rate with respect to Hubble rate and $W_{\Delta L=2}$ denotes the washout coming from $\Delta L =2$ scattering process\footnote[3]{Such processes are explained in \cite{pilaftsis2004resonant}}.

\begin{figure}[!ht]
	\begin{center}
		\includegraphics[width=0.65\textwidth]{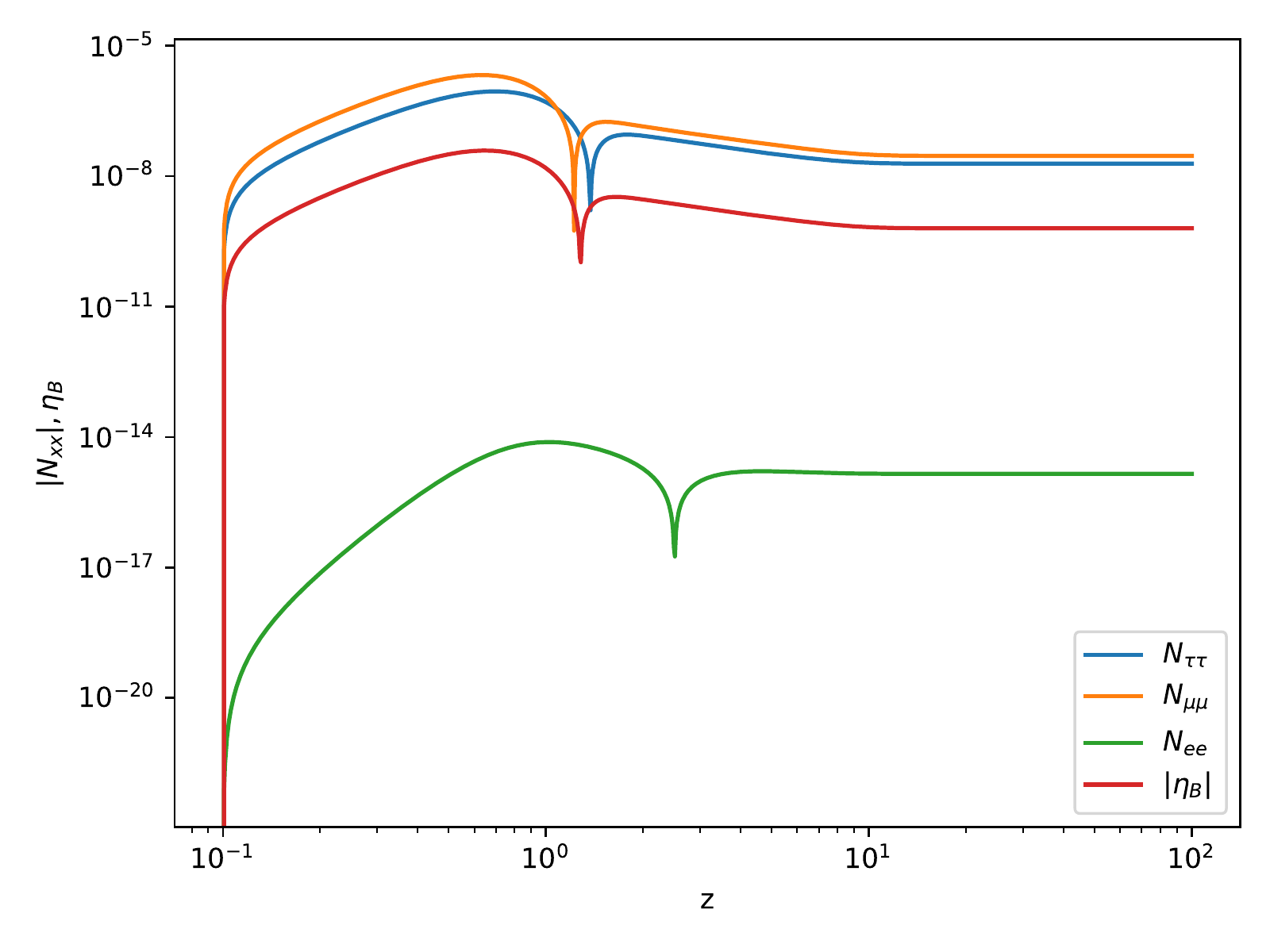}
		\caption{Variation of lepton number density for three flavours and baryon asymmetry parameter, $\eta_B$ as a function of $z$.} 
		\label{fig:bau}
	\end{center}
\end{figure}

\par We take $M_1 = 10\ $TeV and $d = \left(M_2-M_1\right)/M_1 \simeq 10^{-8}$ in order to estimate the value of BAU. We made the calculations related to baryon asymmetry using the ULYSSES package \cite{granelli2021ulysses}. Figure \ref{fig:bau} shows the evolution of three flavoured lepton number density, $N_{\alpha\alpha}$ and baryon asymmetry, $\eta_B$ as a function of $z = \textrm{M}_{N_1}/T$. The asymptotic value suggests that the obtained value of baryon asymmetry is $\lvert\eta_B\rvert \approx 6.3 \times 10^{-10}$. 

\begin{figure}[!ht]
	\begin{center}
		\includegraphics[width=0.6\textwidth]{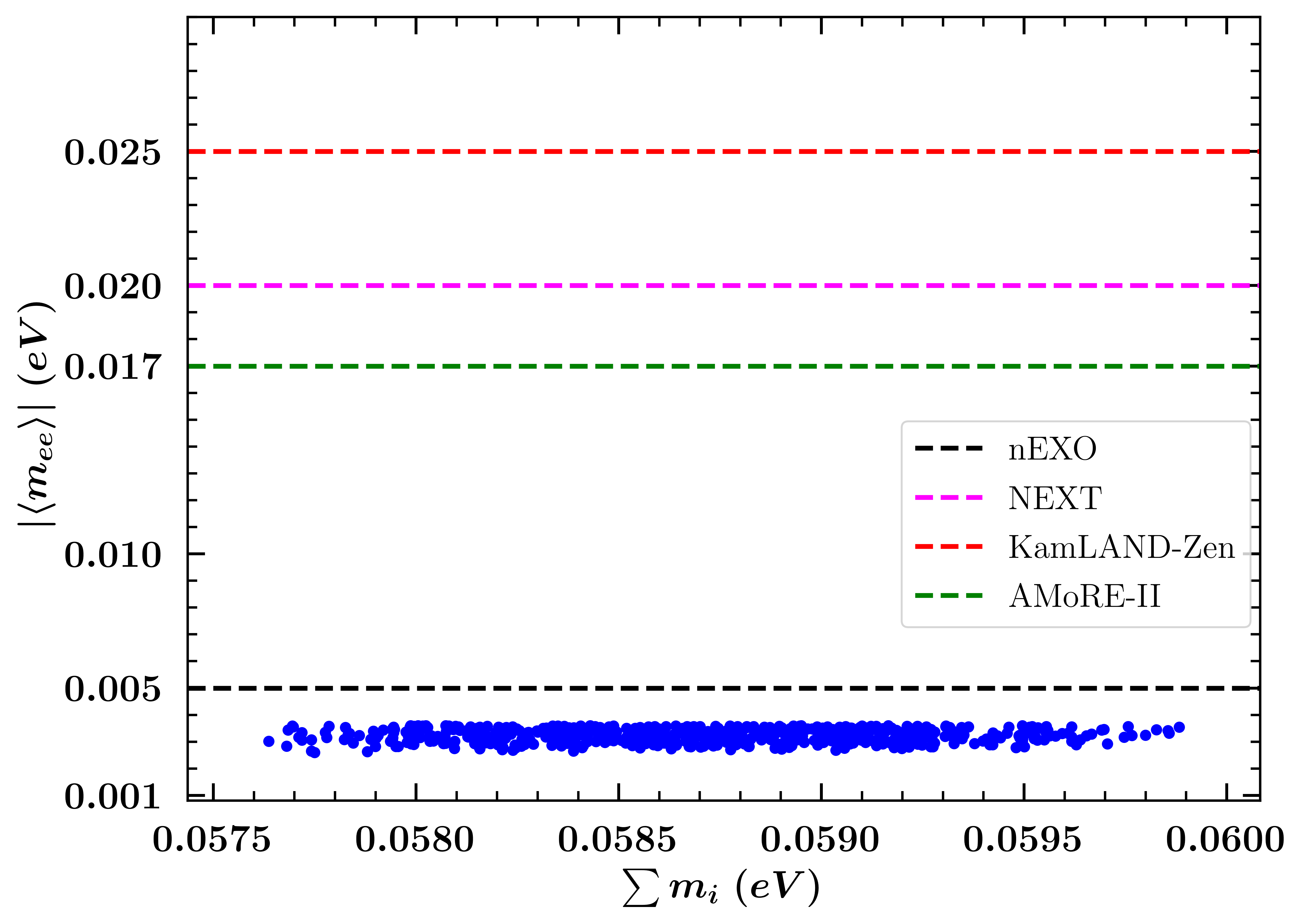}
		\caption{The predicted values of $\lvert \langle m_{ee} \rangle \rvert$ with respect to $\sum m_i$.}
		\label{fig:mee}
	\end{center}
\end{figure}

\par The effective Majorana mass relevant for the neutrinoless double beta decay  ($0\nu\beta\beta$) is the (1, 1) element of the effective neutrino mass matrix (Eq 8), $\lvert \langle m_{ee} \rangle \rvert = \lvert (m_\nu)_{11}\rvert$. Figure \ref{fig:mee} shows the predicted values of $\lvert \langle m_{ee} \rangle \rvert$ with respect to $\sum m_i$ for the allowed region of parameter space. We have also shown the sensitivity reach of some experiments such as nEXO \cite{licciardi2017sensitivity}, KamLAND-Zen \cite{obara2017status}, NEXT \cite{gomez2014present}, AMoRE-II \cite{bhang2012amore}. It shows that $\lvert \langle m_{ee} \rangle \rvert$ ranges from 2.6 meV to 3.6 meV and probing such small parameters by ($0\nu\beta\beta$) experiments would be quite difficult.

\section{Conclusion}
We have explored the S$_4$ symmetric flavour model in the context of minimal Type-I seesaw mechanism leading to TM$_1$ mixing pattern in the leptonic sector. In order to achieve TM$_1$ mixing, we extended the scalar sector further by adding a flavon $\psi$ and its $vev$ is chosen such that it follows the orthogonality conditions (i.e., $\langle \psi \rangle \cdot\langle \phi_l \rangle$ and $\langle \psi \rangle \cdot\langle \phi_\nu \rangle$). The resulting effective neutrino mass matrix predicts NH for masses of the neutrinos and $0.0576 \ \textrm{eV} < \sum m_i < 0.0599 \ \textrm{eV}$. An allowed region for the model parameters is derived numerically such that the predictions on the mixing angles, CP phase, and mass squared differences lie within the 3$\sigma$ bound of current oscillation data. Among various points within the 6-dimensional parameter space, the best-fit value is obtained through chi-squared analysis. Using the obtained parameter space, we evaluated the effective Majorana neutrino mass, $\lvert \langle m_{ee} \rangle\rvert$ and we found that it is relatively small, and difficult to probe at the $0\nu\beta\beta$ experiments.\par
Furthermore, we investigated baryogenesis via flavoured resonant leptogenesis. The right-handed neutrinos are degenerate at dimension 5 level, and hence a tiny splitting was generated by including higher dimension term. We have taken the splitting parameter, $d \simeq 10^{-8}$ and thus, obtained a non-zero, resonantly enhanced CP asymmetry from the out-of-equilibrium decay of right-handed Majorana neutrinos. The analysis of the evolution of particles and asymmetry is done by solving the Boltzmann equations. Here, the best-fit values for the model parameters is considered as inputs, and the Boltzmann equations are solved numerically to estimate baryon asymmetry. It was found that the predicted baryon asymmetry comes out to be $\lvert\eta_B\rvert \approx 6.3 \times 10^{-10}$.

\section*{Acknowledgement}
BT acknowledges the Department of Science and Technology (DST), Government of India for INSPIRE Fellowship vide Grant No. DST/INSPIRE/2018/IF180588. The research of NKF is funded by DST-SERB, India under Grant No. EMR/2015/001683.

\section*{Appendix}
\appendix
\section{S$_4$ group}
The irreducible representations of S$_4$ follow the following Kronecker products,
\begin{equation*}
\begin{split}
    & 1_1 \otimes \eta = \eta, \quad 1_2\otimes1_2 = 1, \quad 1_2 \otimes 2 = 2, \quad 1_2 \otimes3_1 = 3_2,\quad 1_2 \otimes 3_2 = 3_1\\ &
    2 \ \otimes \ 2 = 1_1\oplus1_2\oplus2, \quad 2\otimes3_1=2\otimes3_2=3_1\oplus3_2,\\ &
    3_1\otimes3_1=3_2\otimes3_2=1_1\oplus2\oplus3_1\oplus3_2, \quad 3_1\otimes3_2=1_2\oplus2\oplus3_1\oplus3_2
\end{split}   
\end{equation*}
Now, we write the Clebsch-Gordon coefficients in particular basis \cite{Bazzocchi2009FermionMA}\\
\noindent For 1-dimensional representations:  
\begin{align*}
1_1 \otimes \eta & = \eta \otimes 1_1 =\eta\\
1_2 \otimes 1_2 & = 1_1 \sim \alpha \beta\\
1_2 \otimes 2 & = 2 \sim 
\begin{pmatrix}
	\alpha\beta_1\\
	-\alpha\beta_2
\end{pmatrix}\\
1_2 \otimes 3_1 & = 3_2 \sim
\begin{pmatrix}
	\alpha\beta_1\\
	\alpha\beta_2\\
	\alpha\beta_3
\end{pmatrix}\\
1_2 \otimes 3_2 & = 3_1 \sim
\begin{pmatrix}
	\alpha\beta_1\\
	\alpha\beta_2\\
	\alpha\beta_3
\end{pmatrix}
\end{align*}

\noindent For 2-dimensional representations:
\begin{align*}
2 \otimes 2 &= 1_1 \oplus 1_2 \oplus 2 &&\textrm{with} 
\begin{cases}
1_1 \sim \alpha_1 \beta_2 + \alpha_2 \beta_1\\
1_2 \sim \alpha_1 \beta_2 - \alpha_2 \beta_1\\
2 \sim
\begin{pmatrix}
\alpha_2 \beta_2\\
\alpha_1 \beta_1
\end{pmatrix}
\end{cases}
\end{align*}
\begin{align*}
2 \otimes 3_1 &= 3_1 \oplus 3_2 &&\textrm{with} 
\begin{cases}
3_1 \sim 
\begin{pmatrix}
	\alpha_1 \beta_2 + \alpha_2 \beta_3\\
	\alpha_1 \beta_3 + \alpha_2 \beta_1\\
	\alpha_1 \beta_1 + \alpha_2 \beta_2
\end{pmatrix}\\
3_2 \sim
\begin{pmatrix}
	\alpha_1 \beta_2 - \alpha_2 \beta_3\\
	\alpha_1 \beta_3 - \alpha_2 \beta_1\\
	\alpha_1 \beta_1 - \alpha_2 \beta_2
\end{pmatrix}
\end{cases}\\
2 \otimes 3_2 &= 3_1 \oplus 3_2 &&\textrm{with} 
\begin{cases}
3_1 \sim 
\begin{pmatrix}
	\alpha_1 \beta_2 - \alpha_2 \beta_3\\
	\alpha_1 \beta_3 - \alpha_2 \beta_1\\
	\alpha_1 \beta_1 - \alpha_2 \beta_2
\end{pmatrix}\\
3_2 \sim
\begin{pmatrix}
	\alpha_1 \beta_2 + \alpha_2 \beta_3\\
	\alpha_1 \beta_3 + \alpha_2 \beta_1\\
	\alpha_1 \beta_1 + \alpha_2 \beta_2
\end{pmatrix}
\end{cases}
\end{align*}
\noindent For 3-dimensional representations:
\begin{align*}
3_1 \otimes 3_1 &= 3_2 \otimes 3_2 =1_1 \oplus 2 \oplus 3_1 \oplus 3_2 &&\textrm{with} 
\begin{cases}
1_1 \sim \alpha_1 \beta_1 + \alpha_2 \beta_3 + \alpha_3\beta_2\\
2 \sim
\begin{pmatrix}
\alpha_2 \beta_2 + \alpha_1\beta_3 + \alpha_3\beta_1\\
\alpha_3 \beta_3 + \alpha_1\beta_2 + \alpha_2\beta_1
\end{pmatrix}\\
3_1 \sim
\begin{pmatrix}
	2\alpha_1 \beta_1 - \alpha_2 \beta_3 - \alpha_3 \beta_2\\
	2\alpha_3 \beta_3 - \alpha_1 \beta_2 - \alpha_2 \beta_1\\
	2\alpha_2 \beta_2 - \alpha_1 \beta_3 - \alpha_3 \beta_1
\end{pmatrix}\\
3_2 \sim
\begin{pmatrix}
	\alpha_2\beta_3 - \alpha_3\beta_2\\
	\alpha_1\beta_2 - \alpha_2\beta_1\\
	\alpha_3\beta_1 - \alpha_1\beta_3
\end{pmatrix}
\end{cases}
\end{align*}
\begin{align*}
3_1 \otimes 3_2 &= 1_2 \oplus 2 \oplus 3_1 \oplus 3_2 &&\textrm{with} 
\begin{cases}
1_2 \sim \alpha_1 \beta_1 + \alpha_2 \beta_3 + \alpha_3\beta_2\\
2 \sim
\begin{pmatrix}
\alpha_2 \beta_2 + \alpha_1\beta_3 + \alpha_3\beta_1\\
-\alpha_3 \beta_3 - \alpha_1\beta_2 - \alpha_2\beta_1
\end{pmatrix}\\
3_1 \sim
\begin{pmatrix}
	\alpha_2\beta_3 - \alpha_3\beta_2\\
	\alpha_1\beta_2 - \alpha_2\beta_1\\
	\alpha_3\beta_1 - \alpha_1\beta_3
\end{pmatrix}\\
3_2 \sim
\begin{pmatrix}
	2\alpha_1 \beta_1 - \alpha_2 \beta_3 - \alpha_3 \beta_2\\
	2\alpha_3 \beta_3 - \alpha_1 \beta_2 - \alpha_2 \beta_1\\
	2\alpha_2 \beta_2 - \alpha_1 \beta_3 - \alpha_3 \beta_1
\end{pmatrix}
\end{cases}
\end{align*}
where $\alpha_i$ and $\beta_i$ denotes the elements of the first and second elements, respectively.

\bibliographystyle{unsrt}
\bibliography{references}

\end{document}